\documentclass[aps,pre,floats,twocolumn,superscriptaddress,reprint]{revtex4-1}

\usepackage{graphicx,epsfig}
\usepackage{graphics,dcolumn,bm,epic, eepic,fleqn,float}
\usepackage{amssymb,amsmath,amsfonts,multirow,rotate,color}

\usepackage[table]{xcolor}
\usepackage{flushend}

\DeclareMathOperator*{\argmin}{argmin}
\begin{document}

\title{Quantifying ethnic segregation in cities through random walks}

\author{Sandro Sousa}
\affiliation{School of Mathematical Sciences, Queen Mary University
  of London, London E1 4NS, United Kingdom}

\author{Vincenzo Nicosia}
\affiliation{School of Mathematical Sciences, Queen Mary University
  of London, London E1 4NS, United Kingdom}

\begin{abstract}
  Socioeconomic segregation is considered one of the main factors
  behind the emergence of large-scale inequalities in urban areas, and
  its characterisation is an active area of research in urban
  studies. There are currently many available measures of spatial
  segregation, but almost all of them either depend in non-trivial
  ways on the scale and size of the system under study, or mostly
  neglect the importance of large-scale spatial correlations, or
  depend on parameters which make it hard to compare different systems
  on equal grounds.  We propose here two non-parametric measures of
  spatial variance and local spatial diversity, based on the
  statistical properties of the trajectories of random walks on
  graphs.  We show that these two quantities provide a consistent and
  intuitive estimation of segregation of synthetic spatial patterns,
  and we use them to analyse and compare the ethnic segregation of
  large metropolitan areas in the US and the UK. The results confirm
  that the spatial variance and local diversity as measured through
  simple diffusion on graphs provides meaningful insights about the
  spatial organisation of ethnicities across a city, and allows us to
  efficiently compare the ethnic segregation of urban areas across the
  world irrespective of their size, shape, or peculiar microscopic
  characteristics.
\end{abstract}

\maketitle

Spatial heterogeneity is a characteristic aspect of a variety of
complex systems, from urban areas to
ecosystems~\cite{Gelfand2010,barthelemy2016urban}, and the presence of
non-trivial spatial patterns in the organisation of such systems has a
substantial impact on their functioning and
dynamics~\cite{batty2017new}. This is the main reason why the
quantitative characterisation of complex spatial patterns has received
much attention in different fields, from urban studies to biology,
from geography to economics, from transportation to
engineering~\cite{pickett1995landscape,pickett1995landscape,irwin2007evolution,randon2018urban,brelsford2017heterogeneity}.

A particularly pressing problem in this field is the quantification of
spatial segregation, i.e., the tendency of the units of a system to
form uniform agglomerates around closely-located areas
(neighbourhoods, census tracts, wards, etc.). A typical example is
that of segregation of urban areas by socio-economic indicators,
including ethnicity, income, education or religion, which is known to
be associated with urban wealth, security, and
liveability~\cite{barthelemy2016urban,batty2017new}. The standard
approach in this case is to devise measures of how the local density
and heterogeneity of the property under study, as obtained from census
data at a given scale, compares with the distribution at the system
level, under the assumption that in a non-segregated system the local
distribution of, say, ethnicity would closely mirror the overall
distribution at the city
level~\cite{sakoda1981generalized,reardon2002multi,reardon2004measures,feitosa2007global,sullivan2007surface,reardon2008geographic,wong2011measuring}.

There is general agreement about the fact that spatial segregation is
a multifaceted characteristic of a system. Indeed, the literature
distinguishes different dimensions of the
phenomenon~\cite{massey1988dimensions,reardon2004measures}, namely
spatial exposure/isolation, which measures the extent to which the
members of one group are in close contact with members of another
group due to their placement in space, and spatial
evenness/clustering, which quantifies how uniformly groups are
distributed in space. Despite this framework is extensively used when
referring to spatial segregation patterns, there is no consensus on
how these aspects of spatial segregation should be quantified, or on
how to compare the levels of segregation in different urban
systems~\cite{Gelfand2010,barthelemy2016urban,batty2017new}. In fact,
quantifying spatial segregation is still problematic, mainly because
most of the measures proposed in the literature depend on the scale at
which neighbourhoods are defined, on the granularity of the census
data available, or on the presence of free
parameters~\cite{bojanowski2014measuring,rodriguez2016overview}.  A
growing body of literature has recently started examining urban
segregation within the paradigm of network
science~\cite{barthelemy2011spatial,barthelemy2016urban,barthelemy2018morph},
which consists in analysing one or more graphs associated to an urban
system ---e.g., census tract adjacency, urban transportation,
commuting, etc.--- and deriving descriptive statistics from the
ratios of within-group and between-group connectivity in those
graphs~\cite{gupta1989networks,fershtman1997cohesive,newman2003mixing,echenique2007measure,golub2012how,ballester2014random}.

Here we propose a principled framework to quantify the level of
segregation of a spatial system and to compare the segregation of
different systems, based on the statistics of random walks on
graphs~\cite{noh2004random,fronczak2009biased,bonaventura2014characteristic,masuda2017random,ballester2014random,olteanu2019segregation}.
We consider the symbolic time series of node properties generated by
the trajectories of a random walker through the graph, and we analyse
the spatial distribution of the Class Coverage Time (CCT), that is the
expected number of steps required by a random walk to visit a certain
fraction of all the classes present in the system, when starting from
a generic node.  We purposely avoid here the challenge of directly
defining segregation, assuming that the absence of segregation is
indicated by the concordance of the statistics of coverage time with
those observed in an appropriate null-model. We use this framework to
quantify the ethnic segregation of urban systems in the US and the UK,
and we show that the distribution of class coverage times provides
quite useful insight on the microscopic, meso-scopic, and macro-scopic
organisation of population throughout a city.

\begin{figure}[!t]
  \begin{center}
    \includegraphics[width=3.4in]{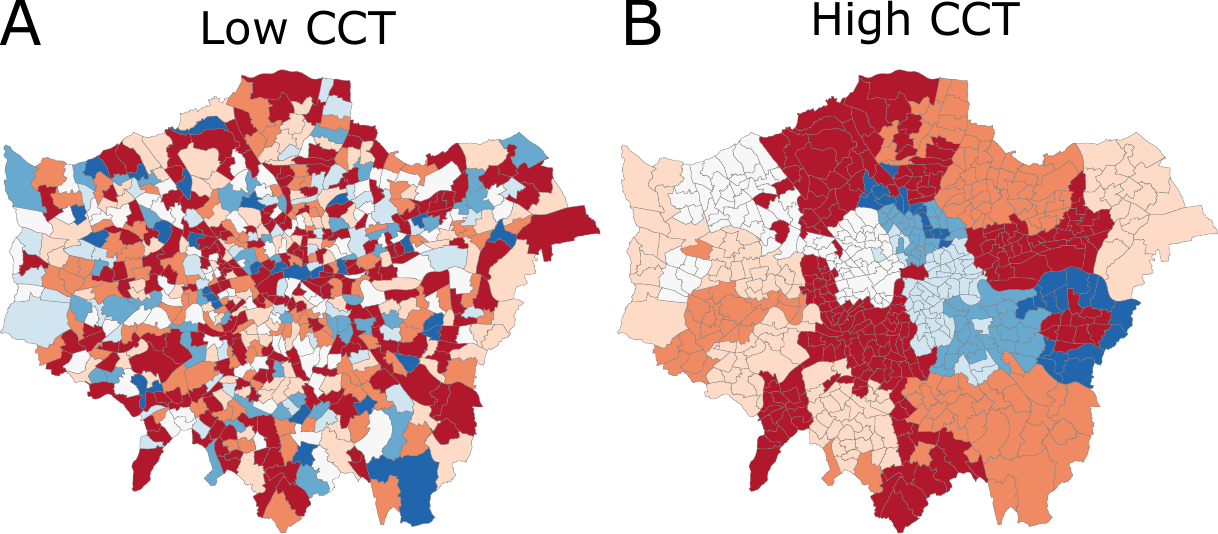}
  \end{center}
  \caption{Fictitious maps of Greater London with 7 ethnicities
    associated to each ward. In panel (A) the ethnicities are
    distributed uniformly at random across the city, to simulate a
    maximally homogeneous and unsegregated pattern. In this case, a
    random walker starting from any ward will get in touch with all
    the available ethnicities within a relatively small number of
    steps. In panel (B) we imposed an artificial and substantial
    clustering of ethnicities, In this case, a walker starting in the
    middle of a cluster will need a lot more time to visits all the
    other ethnicities. This observation leads to the idea of using the
    statistics of Class Coverage Time to quantify the level of
    segregation and heterogeneity of an urban area with respect to a
    given variable of interest.}
  \label{fig:fig1}
\end{figure}

\section*{Model}
\label{sec:model}

Let us consider a spatial graph~\cite{barthelemy2011spatial} $G(V,E)$
consisting of $N=|V|$ nodes and $K=|E|$ edges, and assume that each
node $i$ is associated to a certain variable of interest $x_i$, which
can in principle be either a scalar value, e.g., the average income of
people living in the area represented by node $i$, or vectorial, e.g.,
the ethnic distribution at node $i$. We are interested in
characterising the spatial distribution of $x_i$, that is, to which
extent nodes being close to each other in the graph also have similar
values of $x_i$ or form homogeneous clusters. In the specific case of
urban segregation, we actually want to quantify how homogeneous is the
distribution of $\Gamma$ distinct groups across a city, where the
groups can represent ethnicities, income classes, education levels,
etc. In this case, the variable of interest at each node $i$ is the
vector $x_i = \{m_{i,1}, m_{i,2}, \ldots, m_{i,\Gamma}\}$, where
$m_{i,\alpha}$ is equal to the number of citizens of class $\alpha$
living in the census tract associated to node $i$.

Moving from the observation that uniform random walks on a graph
preserve a lot of information about the structure of the
graph~\cite{Rosvall2008,kantelhardt2002multifractal,nicosia2014charac},
we propose to quantify the heterogeneity of the spatial distribution
of $x_i$ by means of the temporal statistics of the symbolic dynamics
$\{\varphi_{i_0}, \varphi_{i_1}, \varphi_{i_2}, \ldots\}$ associated
to the generic trajectory $\{i_0, i_1, i_2,\ldots\}$ of a uniform
random walk on $G$~\cite{masuda2017random}, where $\varphi_{i_t}$ is
an appropriately-chosen function of $x_{i_t}$. It is worth stressing
that in general $\varphi_{i_t}$ can be constructed in many different
ways, and could depend not only on the specific quantity $x_i$ we are
interested in, but also on the actual time $t$ at which a walker
visits node $i$.

As a simple example, we show in Fig.~\ref{fig:fig1} the planar graphs
associated to the map of wards in Greater London (UK). Here, nodes
represent wards and two nodes are connected by an edge if the
corresponding wards are bordering each other. We associate to each
node $i$ the vectorial variable $x_i=\{m_{i,1}, m_{i,2}, \ldots,
m_{i,\Gamma}\}$, that is the (fictitious) ethnicity distribution of
the residents at $i$. In this case, we set $\varphi_{i}$ equal to the
most abundant ethnicity at $i$, and we coloured each node in the
figure accordingly. In Fig.~\ref{fig:fig1}A the ethnicities are
distributed uniformly at random across the city, simulating a
maximally homogeneous and unsegregated pattern, while in
Fig.~\ref{fig:fig1}B we imposed an artificial clustering of
ethnicities in neighbouring areas, to simulate an extremely segregated
scenario. Note that a random walker starting from any of the areas in
panel A will in general require a small amount of time to encounter
all the classes (ethnicities), since each class is reachable from any
ward within a few hops. Conversely, a walker starting inside one of
the large clusters in panel B will require a considerably larger
amount of time to visit another class, and a comparatively much larger
amount of time to visit all the classes in the city. This suggests
that the number of time steps needed to a random walker to visit all
the ethnicities in a urban area actually contains very useful
information about the spatial organisation of ethnicities across the
system.

Starting from this observation, we propose here to quantify the level
of segregation of an urban area with respect to a categorical variable
that the $\Gamma$ ethnicities present in an urban area by means of the
Class Coverage Time (CCT) of a random walk on the corresponding
graph. This is the number of steps needed by a walker started at a
generic node $i_0$ to visit a prescribed fraction $c$ of all the
$\Gamma$ classes. If classes are distributed uniformly across the
city, the class coverage time will not depend heavily on the starting
node $i_0$. Conversely, if classes tend to form homogeneous groups and
clusters, then the amount of steps needed to visit a given fraction
$c$ of all the ethnic groups present in the city will actually depend
on the shape, size, and depth of the cluster from which the walker
starts and of the other clusters present in the system. In general,
higher heterogeneity in the spatial distribution of class coverage
time will correspond to higher spatial constraints and signal the
presence of segregation.

More formally, let us consider a random walk that starts from a node
$i$ and visits the sequence of nodes
$\{i=i_0,i_1,i_2,\ldots,i_t,\ldots\}$ at subsequent discrete time
steps $t=0,1,2,\ldots,t, \ldots$. We call $\mathcal{W}_{i}(t)$ the
fraction of distinct classes encountered by the walker up to time $t$
when it started from node $i$ at time $0$, and we compute the average
over $R$ independent realisations of the walk:
\begin{equation}
  \overline{\mathcal{W}}_{i}(t) = \frac{1}{R} \sum_1^{R}
  \mathcal{W}_{i}(t)
\end{equation}
We define the Class Coverage Time ($CCT$) of node $i$ at level $c$ as
the expected number of steps after which the walker has encountered a
fraction $c$ of the $\Gamma$ classes for the first time, that is:
\begin{equation}
  C_{i}(c) = \argmin\limits_t \left\lbrace
  \overline{\mathcal{W}}_{i}(t) = c \right\rbrace
  \label{eq:cct}
\end{equation}
We characterise the distribution of Class Coverage Time of a given
system by looking at its mean:
\begin{equation}
  \mu(c) = \frac{1}{N} \sum_{i=1}^{N} C_i(c),
  \label{eq:mu}
\end{equation}
its coefficient of variation:
\begin{equation}
  \sigma(c) = \sqrt{Var(C_i(c))} / \mu(c),
  \label{eq:sigma}
\end{equation}
and at the level of local spatial diversity, as measured by:
\begin{equation}
  \varrho(c) = \frac{1}{K} \sum_{i=1}^{N} \sum_{j<i} a_{ij} \vert
  C_i(c) - C_j(c) \vert.
  \label{eq:rho}
\end{equation}
In general, larger values of $\mu(c)$ indicate a more heterogeneous
distribution of classes through the system. Similarly, larger values
of $\sigma(c)$ correspond to a larger dependence of CCT on the
starting point, which suggest that $\sigma(c)$ is measuring the
overall spatial variance of class coverage times. Finally, larger
values of $\varrho(c)$ indicate that neighbouring nodes have very
different class coverage times, signalling the presence of local
spatial diversity.

As we will show in the following, all these three measures are somehow
affected by the relative abundance of each class and by the size of
the graph. For this reason, we will consider the average deviation of
each quantity from the corresponding quantity measured in a
null-model:
\begin{equation}
  \Delta\sigma = \int_{0}^{1} {\rm d}c \> \left\vert\sigma(c) -
  \sigma(c)^{\text{null}}\right\vert
  \label{eq:normsigma}
\end{equation}
and
\begin{equation}
  \Delta\varrho = \int_{0}^{1} {\rm d}c \> \left\vert\varrho(c) -
  \varrho(c)^{\text{null}}\right\vert
  \label{eq:normrho}
\end{equation}
The null model we consider consists of the same graph $G$ as the
original system, where the ethnicities distributions $\{x_i\}$ have
been reassigned to nodes uniformly at random, thus maintaining the
relative abundance of classes and the way they tend to be distributed
in a single area while destroying any existing spatial organisation of
classes~\cite{barthelemy2016urban} (see Additional Methods for
details). By comparing $\sigma(c)$ and $\varrho(c)$ with respect to
the null-model we obtain a principled way to compare spatial systems
with different number of classes and characterised by distinct shapes
and scales. In the following, we call $\Delta \sigma$ ``spatial
variance'' and $\Delta \varrho$ ``spatial diversity'', for obvious
reasons.

\begin{figure}[!t]
\centering
\includegraphics[width=3.4in]{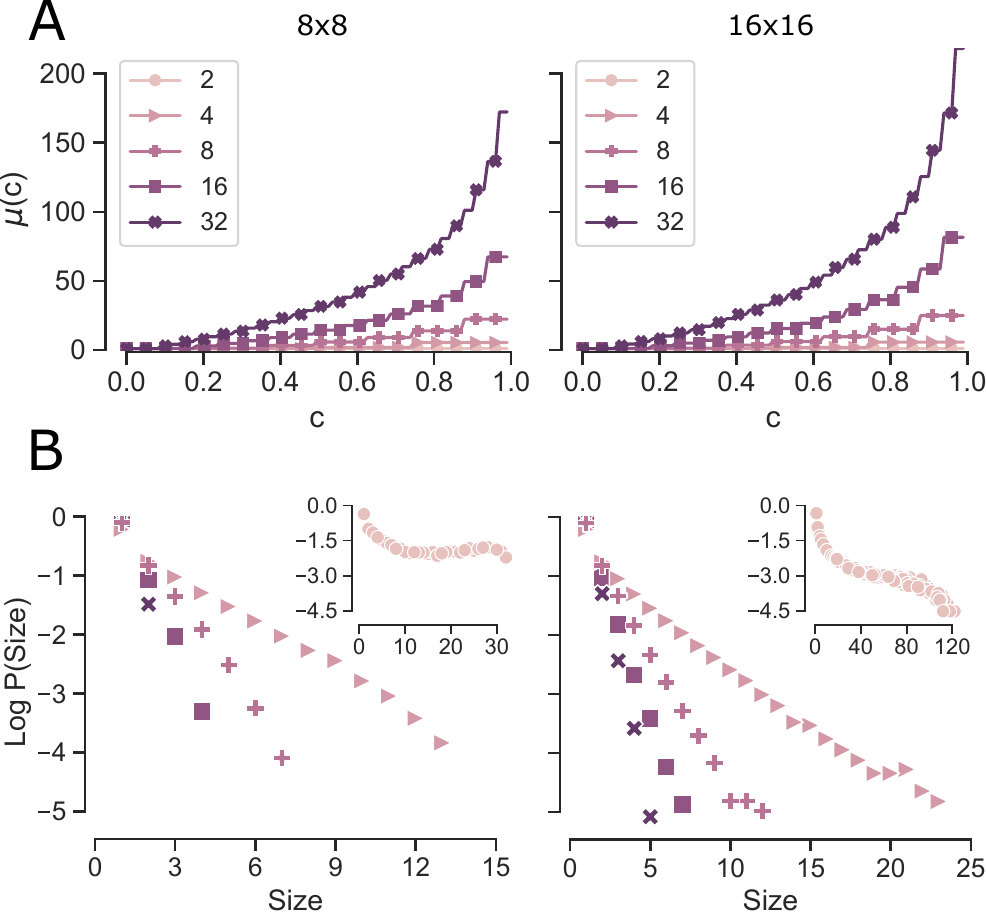}
\caption{Effect of graph size and number of classes on class coverage
  times in synthetic system where $\Gamma$ classes are distributed
  uniformly at random across the nodes. (A) Mean class coverage time
  $\mu(c)$ as a function of $c$ on a torus with 64 (left) and 256
  (right) cells. Cover times over 32 classes are significantly larger
  than over 2 classes. (B) Size distribution of uniform clusters
  formed by adjacent nodes of the same classes. Larger graph allow
  for bigger clusters to emerge, for any number of classes
  $\Gamma$. The legend in (A) marks the corresponding distributions in
  (B). The distributions of cluster sizes for $\Gamma=2$ are reported
  in the insets.}
\label{fig:scale}
\end{figure}

\begin{figure*}[!t]
\centering
\includegraphics[width=7in]{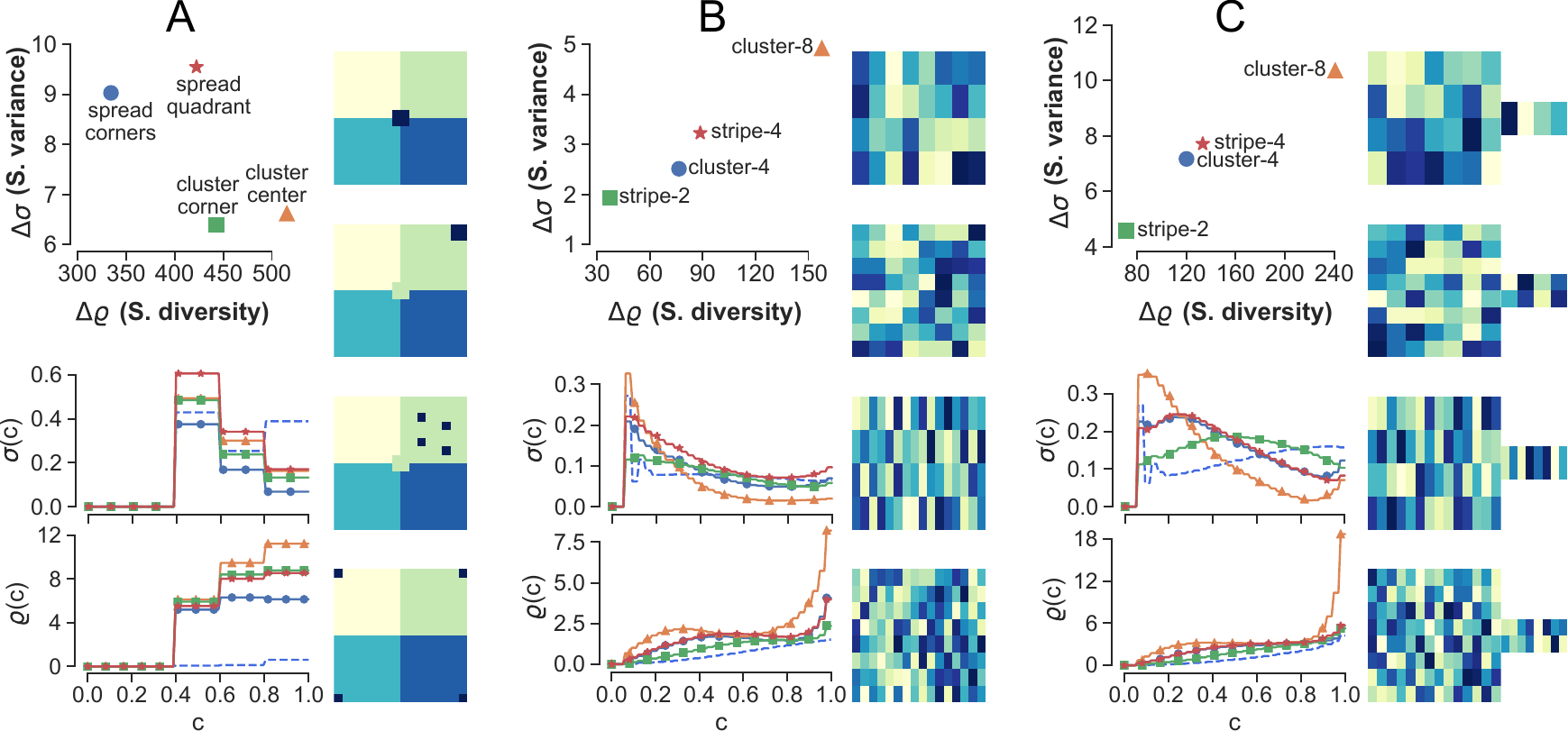}
\caption{Dependence of class coverage times on number of colours, size
  and shape of domain, and size and shape of homogeneous clusters in
  synthetic 2-dimensional lattices (each colour indicates a different
  class). (A) The nodes are divided in 4 classes of 63 cells placed at
  quadrants, while the four nodes in the fifth class are arranged,
  from top to bottom, as a central cluster (\textit{cluster-center}),
  a cluster in a corner (\textit{cluster-corner}), spread in one of
  the quadrants (\textit{spread-quadrant}), and on each of the corners
  of the lattice (\textit{spread-corners}). (B) 32 classes are
  associated uniformly at random (from top to bottom) to clusters of 8
  or 4 cells and to stripes of 4 or 2 cells, on a $16\times 16$
  lattice. (C) Same patterns as in (B), but on a lattice with a
  lateral appendix. We also report in each panel the corresponding
  profiles of $\sigma(c)$ and $\varrho(c)$, and their values in the
  corresponding null model (blue dashed lines). The distinct spatial
  constraints are consistently discriminated by the proposed measured
  of spatial variance and spatial diversity.}
\label{fig:examples}
\end{figure*}

\section*{Simple geometries and synthetic class distributions}

In this section we explore the behaviour of Class Coverage Time in
simple planar lattices. We start from the simple case of
two-dimensional lattices with periodic boundary conditions (tori)
where each node is associated to one of the $\Gamma$ available classes
with uniform probability. In Fig.~\ref{fig:scale}A we report the plot
of $\mu(c)$ as a function of the fraction $c$ of classes reached by
the walker on $8\times 8$ (left panel) and $16\times 16$ lattices
(right panel), for $\Gamma=\{2,4,8,16,32\}$.  As expected, $\mu(c)$ is
a non-linear increasing function of $c$, meaning that reaching a
higher fraction of the classes becomes harder and harder. Moreover,
$\mu(c)$ is also an increasing function of $\Gamma$ for a fixed
fraction $c$, meaning that configurations with more classes typically
exhibit larger coverage times, as expected.  By comparing the two
panels it becomes clear that covering a given percentage $c$ of
classes requires comparatively more time on a larger lattice.

These results are somehow expected, and are due to the fact that even
when classes are distributed uniformly at random, local clusters of
nodes of the same the same class eventually emerge. In particular,
smaller values of $\Gamma$ and larger lattice sizes have a higher
probability of producing larger clusters, which effectively contribute
to reducing the probability that the random walk finds a new class at
each time step. Indeed, a walker that enters a large homogeneous
cluster will in general require more time to visit other
classes. These observations are confirmed by Fig.~\ref{fig:scale}B,
where we report the distributions of cluster sizes for different
values of $\Gamma$ (the case $\Gamma=2$ is in the insets). The larger
values of $\mu(c)$ observed in $16\times 16$ graph in
Fig.~\ref{fig:scale}A can indeed be associated to the presence of
bigger uniform clusters.

To study the effect of domain shape and class placement, we considered
finite lattices with pre-assigned class distributions organised in
specific patterns (the corresponding distributions of Class Coverage
Time are reported in Supplementary Figure S1). In
Fig.~\ref{fig:examples}A we show four arrangements of five classes in
a square lattice. In each arrangement, four of the classes contain
$N/4 - 1$ nodes, and form homogeneous clusters which occupy a quadrant
each. The four nodes in the fifth class, instead, (i) either form a
single cluster in the centre (top panel), or (ii) in one corner of the
lattice (second panel from the top), or (iii) are scattered within
another cluster (third panel from the top), or (iv) placed at the four
corners (bottom panel). Notice that all these four patterns are
associated to the same null-model, since the relative abundances of
the five classes are kept constant. Indeed, pattern (i) and (ii)
exhibit the largest values of local spatial diversity and the smallest
values of spatial variance. On the contrary, pattern (iii) and pattern
(iv) have smaller values of local spatial diversity and larger values
of spatial variance, mainly due to the fact that the nodes in the
smallest cluster are on average farther away from all the others.

\begin{figure*}[!t]
\centering
\includegraphics[width=7in]{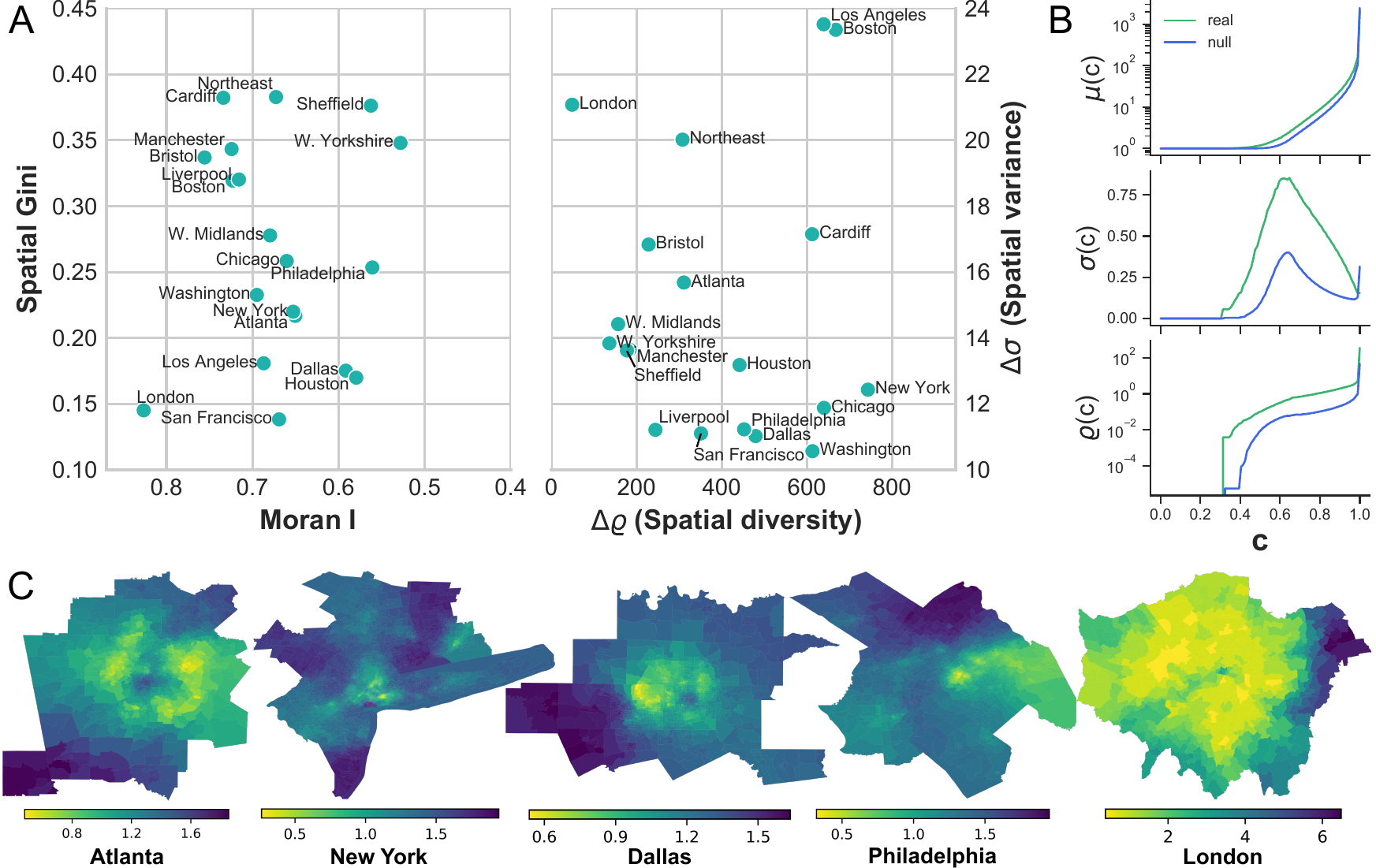}
\caption{Ethnic segregation in urban systems. (A) Spatial variance and
  spatial diversity of metropolitan areas in the US and UK compared to
  their Spatial Gini coefficient and Moran I. For both ranks smaller
  is better, except Moran I where large values indicate higher
  correlation (x-axis inverted for better readability).  (B) Examples
  of the class coverage time distributions for London where the mean
  coverage time $\mu(c)$, standard deviation $\sigma(c)$ and the
  spatial correlation $\varrho(c)$ are plotted as a function of
  fractions $c$ of classes.  (C) The maps of the normalised class
  coverage time $C_i(c)/C_i(c)^{\text{null}}$ for $c=0.7$ provide
  detailed insights about the structure of segregation at
  neighbourhood level.}
\label{fig:main}
\end{figure*}

To investigate the role of the size and shape of local clusters, in
Fig.~(\ref{fig:examples})B we considered four different random tilings
of the same $16\times 16$ two-dimensional lattice in 32 classes,
respectively organised (from top to bottom) in (i) 32 rectangular
clusters of size $2\times 4$ (cluster-8), (ii) 64 square clusters of
size $2\times 2$ (cluster-4), 64 rectangular clusters of size $1\times
4$ (stripe-4) and (iv) 128 rectangular clusters of size $1\times 2$
(stripe-2).  Notice that configuration (i) (cluster-8) corresponds to
the largest possible value of spatial diversity and local spatial
variance. On the other hand, the case of stripe-2 clusters (iv), which
is the most similar to the null-model, yields the smallest values of
spatial diversity and spatial variance, as expected. The relative
positions of intermediate configurations (ii) and (iii) in the $\Delta
\varrho/\Delta\sigma$ plane can be explained by the fact that a tiling
with square clusters provides comparatively lower values of spatial
heterogeneity and spatial variance than clusters of four nodes
arranged in a line, mainly because a square of size $N$ has a smaller
perimeter than a rectangle of the same area, hence more neighbours.

Finally, we show that $\Delta \sigma$ and $\Delta \varrho$ are also
affected by the actual shape of the domain, by considering the 2D
lattice with a lateral appendix in Fig.~(\ref{fig:examples})C. Indeed,
for the same number of classes and the same cluster shapes and sizes
as in Fig.~(\ref{fig:examples})B, the arrangements in
Fig.~(\ref{fig:examples})C correspond to much larger values of spatial
variance and local spatial diversity, mainly due to the fact that the
walkers started at the nodes belonging to the lateral appendix will
require a much larger amount of time to visit a certain fraction of
the classes than the walkers started at nodes in the bulk (see
Supplementary Figure S1 for additional details).

\section*{Ethnic residential segregation in the US and UK}

In this section we show that spatial variance ($\Delta \sigma$) and
local spatial diversity ($\Delta \varrho$) can be effectively used to
compare the ethnic residential segregation of metropolitan areas in a
quantitative way. We used geo--referenced census data for metropolitan
areas in the US and the UK, and for each urban area we constructed the
graph $G$ of physical adjacency between census tracts (US) and wards
(UK) (see Additional Methods for details). Each node is associated to
the distribution of ethnicities in the corresponding area. We computed
the coverage time from each node as in Eq.~(\ref{eq:cct}), and the
corresponding values of spatial variance $\Delta \sigma$ and local
spatial diversity $\Delta \varrho$.  In Fig.~\ref{fig:main}A we report
each urban area in the plane $\Delta \sigma/\Delta \varrho$, and we
compare the results with those obtained using two other classical
measures of segregation extensively used in the literature, namely the
spatial correlation Moran I~\cite{cliff1981} and the spatial Gini
coefficient~\cite{rey2013spatial} (see Additional Methods for details,
Supplementary Table S2 and Supplementary Figures S2-S9 for additional
information). Note that we inverted the x-axis for the Moran I index,
to make the comparison with $\Delta \varrho$ easier.

It is worth noting that some cities which are very close in the
Moran/Gini plane are placed quite differently in the $\Delta
\varrho/\Delta \sigma$ plane, and vice versa. An interesting example
is that of New York and Atlanta, which are similar according to
Moran/Gini and quite far apart from each other according to the
proposed measures. Indeed, if we look at the spatial distribution of
$\widetilde{C}_i(c) = C_i(c) / C_i^{\text{null}}(c)$ in the two cities
(as show in Fig.~(\ref{fig:main})C), the two maps are quite different:
in Atlanta the yellow-ish areas indicating regions with the smaller
values of $C_i(c)$ (i.e., easier access to all the ethnicities) are
those placed around the city centre, while in New York most of the
areas exhibit quite large values of CCT, indicating the presence of a
somehow higher ethnic segregation. This difference is indeed captured
well in the $\Delta \varrho/\Delta \sigma$ plane, according to which
New York has a smaller spatial variance and a larger local diversity
than Atlanta. Conversely, Dallas and Philadelphia, which exhibit a
visually-similar pattern of $\widetilde{C}_i(c)$, are quite far apart
in the Moran/Gini plane but are put very close in the $\Delta
\varrho/\Delta \sigma$ plane.

In Fig.~\ref{fig:main}B we look closely at the behaviour of $\mu(c)$,
$\sigma(c)$ and $\varrho(c)$ in London, which is well-know for being
characterised by strong residential segregation. Indeed, the Moran
index for London is relatively high, and there are some areas of the
city which clearly exhibit substantially larger values of class
coverage time. However, the local spatial diversity is relatively low,
indicating that adjacent regions tend to be organised in clusters
having very similar distribution of ethnicities.

In summary, higher levels of spatial variance of coverage time
$\Delta\sigma \gg 0$ indicate an unbalanced spatial distribution of
ethnicities in the city, meaning that citizens experience large
variations in the time needed to encounter all the other ethnicities
depending on where they live. In contrast, low spatial variance
indicates that on average the spatial distribution of ethnicities is
relatively uniform across the city and individuals living in different
areas are similarly exposed to the ethnicities living in the city. Low
levels of local spatial diversity $\Delta\varrho$ indicate that there
is no significant difference on the coverage time of neighbouring
areas, that is, the constraints driven by spatial shape are not too
important.  When $\Delta\varrho \gg 0$, the differences between
neighbouring nodes is substantial and segregation is influenced by
clusters with similar ethnicity distributions, which indicates the
presence of a preference mechanism, often resulting from social or
economic pressure. Interestingly, Los Angeles and Boston are the two
urban areas whose patterns of class coverage times are farther away
from the corresponding null model.

It is worth noting that all the urban areas analysed in this study
present some level of spatial variance or spatial diversity, despite
the sizes and population of the areas considered span relatively large
intervals. These results are definitely related to the actual
distribution of ethnicities across the urban areas (see Supplementary
Figures S10-S11), and are consistent across different granularity
scales, as confirmed by a detrended fluctuation analysis (DFA) of the
trajectories of random walks on graphs obtained at different spatial
resolutions (see Supplementary Table S1 and Supplementary
Fig. S12).

Despite we have focused exclusively on the characterisation of ethnic
segregation, the methodology introduced here can be used to quantify
the spatial variance and spatial diversity of the distribution of any
categorical variable, including socio-economic indicators like income,
access to services, education level, and so forth. The consistent
behaviour of $\Delta \sigma$ and $\Delta \rho$ across different
scales is indeed a very desirable property of segregation measures,
as also pointed out by both classical and more recent
works~\cite{openshaw1981,chodrow2017structure}. The fact that these
measures are appropriately normalised by comparing with the
corresponding null-models, make them suitable for comparing the
spatial heterogeneity of the same variable in different systems,
irrespective of their peculiar size and shape, of the actual number of
different classes or categories available in each system, and of the
granularity at which spatial information is aggregated.

\section*{Additional Methods}

\subsection*{Data}\label{sec:data}
In this study, we use the UK Office for National Statistics 2011
Census quick statistics tables, which include population estimates
classified by ethnic group. The available territorial divisions are
regions, districts, unitary authorities, MSOAs, LSOAs and OAs in
England and Wales. The households are divided in 250 ethnic groups for
the detailed tables. All data of the 2011 British Census are available
from the Office for National Statistics \cite{ons2011}. The delineations
of the statistical areas are available from the
UK Data Service \cite{ukdata2011}.

For US cities, we use the American Census Bureau’s 2010 Decennial
Census data, which include race/ethnicity of individuals at the Census
Block level. The households are divided in 64 race/ethnic groups for
the detailed tables within the corresponding combined statistical
area. The delineations of the Census blocks are available from the
same agency at the Geography section \cite{us2010tiger}. For metropolitan areas
containing islands as part of the territory, we focused on the largest
surface to avoid the presence of disconnected components on the
corresponding graph.

\subsection*{Null model}
Given a graph $G$ and an assignment of classes to nodes, the null
model consists of randomly reassigning the class distributions of
nodes while preserving the structure of the graph $G$ and the local
population distribution at each node. It is worth noting that the
spatial scale at which the null model is defined is the same of the
system under study, so that problems generated by comparing cities at
distinct scales are reduced to a minimum.

\subsection*{Spurious effects for $c\simeq 1$}

Coverage time distributions for a city will in general depend on the
abundance of the classes and on how they are distributed in space. In
particular, if a class is very rare, i.e., present only in a few
tracts, then the time needed to visit all the classes will effectively
become comparable with the cover time, which is known to scale
exponentially with the size of the graph~\cite{masuda2017random}. We
decided to minimise these spurious effects by removing from the
analysis of CCT the case $c=1$.

\subsection*{Moran I and Spatial Gini Coefficient}

The auto-correlation Moran I coefficient and the spatial Gini
coefficient are respectively comparable to the local spatial diversity
$\varrho(c)$ and to the spatial variance $\sigma(c)$, and they have
the good property of being non-parametric and of having been
extensively used in the literature. We computed these two quantities
using as node variable the Shannon entropy of the ethnicity
distribution in the corresponding area, denoted here by $x_i$. The
Moran I is given by:
\begin{equation}
I = \frac{N}{W} \frac{\sum_i\sum_j w_{ij}(x_i-\langle x\rangle)
(x_j-\langle x\rangle)} {\sum_i (x_i-\langle x\rangle)^{2}}
\end{equation}
and the spatial Gini coefficient is given by:
\begin{equation}
SG = \frac{\sum_{i=1}^{N} \sum_{j=1}^{N} w_{ij}\mid x_i-x_j\mid +
(1-w_{ij})\mid x_i-x_j\mid}{2n^{2}\langle x\rangle}
\end{equation}
where for both measures, $N$ is the number of neighbourhoods and
$\langle x\rangle=\frac{1}{N}\sum_i x_i$ is the mean of the variable
of interest. The spatial weight $w_{ij}$ is defined according to the
adjacency matrix $A$ where $w_{ij}=1$ if two areas are neighbours, and
0 otherwise. The diagonal elements $w_{ii}=0$ as defined in $A$ and
$W$ corresponds to the sum of all weights. Interestingly, the picture
provided by the classical Moran/Gini analysis in Fig.~\ref{fig:main}
does not improve considerably when using other local indicators (see
Supplementary Figures S13-S14).

\section*{Author Contributions}

\noindent
SS and VN devised the study. SS performed the computations. SS and VN
contributed methods, analysed the data, and wrote the manuscript.

\begin{acknowledgments}
  VN acknowledges support from the EPSRC New Investigator Award
  Grant No. EP/S027920/1. This work made use of the MidPLUS cluster,
  EPSRC Grant No. EP/K000128/1.
  \href{http://doi.org/10.5281/zenodo.438045}{doi.org/10.5281/zenodo.438045}.
\end{acknowledgments}


\begin{thebibliography}{10}
\makeatletter

\bibitem{Gelfand2010}
Gelfand AE, Diggle P, Guttorp P, Fuentes M (2010) {\em Handbook of spatial
  statistics}.
\newblock (CRC Press).

\bibitem{barthelemy2016urban}
Barth\'elemy M (2016) {\em The structure and dynamics of cities}.
\newblock (Cambridge University Press).

\bibitem{batty2017new}
Batty M (2017) {\em The new science of cities}.
\newblock (MIT Press).

\bibitem{pickett1995landscape}
Pickett STA, Cadenasso ML (1995) Landscape ecology: Spatial heterogeneity in
  ecological systems.
\newblock {\em Science} 269(5222):331--334.

\bibitem{irwin2007evolution}
Irwin EG, Bockstael NE (2007) The evolution of urban sprawl: Evidence of
  spatial heterogeneity and increasing land fragmentation.
\newblock {\em Proceedings of the National Academy of Sciences}
  104(52):20672--20677.

\bibitem{randon2018urban}
Randon-Furling J, Olteanu M, Lucquiaud A (2018) From urban segregation to
  spatial structure detection.
\newblock {\em Environment and Planning B: Urban Analytics and City Science} p.
  2399808318797129.

\bibitem{brelsford2017heterogeneity}
Brelsford C, Lobo J, Hand J, Bettencourt LM (2017) Heterogeneity and scale of
  sustainable development in cities.
\newblock {\em Proceedings of the National Academy of Sciences}
  114(34):8963--8968.

\bibitem{sakoda1981generalized}
Sakoda JM (1981) A generalized index of dissimilarity.
\newblock {\em Demography} 18(2):245--250.

\bibitem{reardon2002multi}
Reardon SF, Firebaugh G (2002) Measures of multigroup segregation.
\newblock {\em Sociological Methodology} 32(1):33--67.

\bibitem{reardon2004measures}
Reardon SF, O'Sullivan D (2004) Measures of spatial segregation.
\newblock {\em Sociological Methodology} 34(1):121--162.

\bibitem{feitosa2007global}
Feitosa FF, Camara G, Monteiro AMV, Koschitzki T, Silva MPS (2007) Global and
  local spatial indices of urban segregation.
\newblock {\em International Journal of Geographical Information Science}
  21(3):299--323.

\bibitem{sullivan2007surface}
O'Sullivan D, Wong DWS (2007) A surface-based approach to measuring spatial
  segregation.
\newblock {\em Geographical Analysis} 39(2):147--168.

\bibitem{reardon2008geographic}
Reardon SF, et~al. (2008) The geographic scale of metropolitan racial
  segregation.
\newblock {\em Demography} 45(3):489--514.

\bibitem{wong2011measuring}
Wong DWS, Shaw SL (2011) Measuring segregation: an activity space approach.
\newblock {\em Journal of Geographical Systems} 13(2):127--145.

\bibitem{massey1988dimensions}
Massey DS, Denton NA (1988) The dimensions of residential segregation.
\newblock {\em Social forces} 67(2):281--315.

\bibitem{bojanowski2014measuring}
Bojanowski M, Corten R (2014) Measuring segregation in social networks.
\newblock {\em Social Networks} 39:14 -- 32.

\bibitem{rodriguez2016overview}
Rodriguez-Moral A, Vorsatz M (2016) An overview of the measurement of
  segregation: Classical approaches and social network analysis.
\newblock {\em Lecture Notes in Economics and Mathematical Systems}
  683:93--119.

\bibitem{barthelemy2011spatial}
Barth{\'e}lemy M (2011) Spatial networks.
\newblock {\em Physics Reports} 499(1-3):1--101.

\bibitem{barthelemy2018morph}
Barth\'elemy M (2018) {\em Morphogenesis of spatial networks}.
\newblock (Springer).

\bibitem{gupta1989networks}
Gupta S, Anderson RM, May RM (1989) Networks of sexual contacts: implications
  for the pattern of spread of hiv.
\newblock {\em AIDS (London, England)} 3(12):807--817.

\bibitem{fershtman1997cohesive}
Fershtman M (1997) Cohesive group detection in a social network by the
  segregation matrix index.
\newblock {\em Social Networks} 19(3):193--207.

\bibitem{newman2003mixing}
Newman MEJ (2003) Mixing patterns in networks.
\newblock {\em Phys. Rev. E} 67(2):026126.

\bibitem{echenique2007measure}
Echenique F, Fryer RG (2007) A measure of segregation based on social
  interactions.
\newblock {\em The Quarterly Journal of Economics} 122(2):441--485.

\bibitem{golub2012how}
Golub B, Jackson MO (2012) How homophily affects the speed of learning and
  best-response dynamics.
\newblock {\em The Quarterly Journal of Economics} 127(3):1287--1338.

\bibitem{ballester2014random}
Ballester C, Vorsatz M (2014) Random walk-based segregation measures.
\newblock {\em The Review of Economics and Statistics} 96(3):383--401.

\bibitem{noh2004random}
Noh JD, Rieger H (2004) Random walks on complex networks.
\newblock {\em Physical review letters} 92(11):118701.

\bibitem{fronczak2009biased}
Fronczak A, Fronczak P (2009) Biased random walks in complex networks: The role
  of local navigation rules.
\newblock {\em Phys. Rev. E} 80(1):016107.

\bibitem{bonaventura2014characteristic}
Bonaventura M, Nicosia V, Latora V (2014) Characteristic times of biased random
  walks on complex networks.
\newblock {\em Physical Review E} 89(1):012803.

\bibitem{masuda2017random}
Masuda N, Porter MA, Lambiotte R (2017) Random walks and diffusion on networks.
\newblock {\em Physics Reports} 716-717:1 -- 58.

\bibitem{olteanu2019segregation}
Olteanu M, Randon-Furling J, Clark WA (2019) Segregation through the
  multiscalar lens.
\newblock {\em Proceedings of the National Academy of Sciences}
  116(25):12250--12254.

\bibitem{Rosvall2008}
Rosvall M, Bergstrom CT (2008) Maps of random walks on complex networks reveal
  community structure.
\newblock {\em Proceedings of the National Academy of Sciences}
  105(4):1118--1123.

\bibitem{kantelhardt2002multifractal}
Kantelhardt JW, et~al. (2002) Multifractal detrended fluctuation analysis of
  nonstationary time series.
\newblock {\em Physica A: Statistical Mechanics and its Applications} 316(1):87
  -- 114.

\bibitem{nicosia2014charac}
Nicosia V, Domenico MD, Latora V (2014) Characteristic exponents of complex
  networks.
\newblock {\em EPL (Europhysics Letters)} 106(5):58005.

\bibitem{cliff1981}
Cliff AD, Ord JK (1981) {\em Spatial processes: models \& applications}.
\newblock (Taylor \& Francis).

\bibitem{rey2013spatial}
Rey SJ, Smith RJ (2013) A spatial decomposition of the gini coefficient.
\newblock {\em Letters in Spatial and Resource Sciences} 6(2):55--70.

\bibitem{openshaw1981}
Openshaw S (1981) The modifiable areal unit problem.
\newblock {\em Quantitative geography: A British view} pp. 60--69.

\bibitem{chodrow2017structure}
Chodrow PS (2017) Structure and information in spatial segregation.
\newblock {\em Proceedings of the National Academy of Sciences}
  114(44):11591--11596.

\bibitem{ons2011}
{Office for National Statistics} (2010) National records of scotland; northern
  ireland statistics and research agency (2016): 2011 census aggregate data
  (\url{http://dx.doi.org/10.5257/census/aggregate-2011-1}).

\bibitem{ukdata2011}
{UK Data Service Census Support} (2011) Office for national statistics, 2011
  census: Digitised boundary data (england and wales) [computer file]
  (\url{https://borders.ukdataservice.ac.uk/}).

\bibitem{us2010tiger}
{US Census Bureau} (2010) Tiger line shapefiles
  (\url{http://www2.census.gov/geo/tiger/TIGER2010DP1/}).

\end{thebibliography}
\end{document}